\documentclass[preprint,12pt]{elsarticle}
\usepackage[T1]{fontenc}
\usepackage{lmodern}
\usepackage{microtype}
\usepackage{amsmath,amssymb,mathtools}
\usepackage{mathrsfs}
\usepackage{graphicx}
\usepackage{booktabs,tabularx,array,longtable,makecell}
\usepackage{siunitx}
\usepackage{orcidlink}
\usepackage{placeins}
\usepackage{subcaption}
\usepackage{hyperref}
\hypersetup{colorlinks=true,linkcolor=blue,citecolor=blue,urlcolor=blue,
  pdftitle={Topological ModMax-AdS Black Holes: Extremality, Charge Normalization, and Limits of WGC Diagnostics},
  pdfauthor={Saeed Noori Gashti, Behnam Pourhassan}}
\biboptions{numbers,sort&compress}

\newcommand{\dd}{\mathrm{d}}
\newcommand{\cF}{\mathcal{F}}
\newcommand{\cG}{\mathcal{G}}
\newcommand{\cL}{\mathcal{L}}
\newcommand{\cA}{\mathcal{A}}

\newcommand{\AdS}{\mathrm{AdS}}
\newcommand{\ModMax}{\mathrm{ModMax}}

\newcommand{\BF}{\mathrm{BF}}
\newcolumntype{Y}{>{\raggedright\arraybackslash}X}

\begin{document}
\begin{frontmatter}

\title{Topological ModMax--AdS Black Holes: Extremality, Charge Normalization, and Limits of WGC Diagnostics}

\author[du,khazar]{Saeed Noori Gashti\orcidlink{0000-0001-7844-2640}}
\ead{sn.gashti@du.ac.ir}
\ead{saeed.noorigashti70@gmail.com}

\author[du,khazar]{Behnam Pourhassan\orcidlink{0000-0003-1338-7083}}
\ead{b.pourhassan@du.ac.ir}

\author[YU]{Saheb Soroushfar\orcidlink{0000-0003-3151-0532}\corref{cor1}}
\ead{soroush@yu.ac.ir}

\affiliation[du]{organization={School of Physics, Damghan University}, addressline={P. O. Box 3671641167}, city={Damghan}, country={Iran}}
\affiliation[khazar]{organization={Center for Theoretical Physics, Khazar University}, addressline={41 Mehseti Street}, city={Baku}, postcode={AZ1096}, country={Azerbaijan}}
\affiliation[YU]{organization={Department of Physics, Yasouj University}, addressline={75918-74934}, city={Yasouj}, country={Iran}}

\cortext[cor1]{Corresponding author}

\begin{abstract}
We study four-dimensional topological anti-de Sitter black holes sourced by ModMax electrodynamics and its sign-reversed counterpart. Distinguishing the field-strength amplitude from the conserved displacement flux fixes the charge normalization, potential, and first law. We classify the exact extremal roots for spherical, planar, and hyperbolic horizons; in particular, the hyperbolic ModMax branch has no upper-charge discriminant bound and contains a finite-charge zero-mass point. We also derive the unstable spherical null circular orbit and evaluate the local near-horizon AdS$_2$ Breitenlohner--Freedman index for an explicitly specified charged-scalar probe. These are classical diagnostics, not a WGC verdict: the model does not specify a charged-particle spectrum, an ultraviolet completion, or the global perturbation problem. The distance and de Sitter conjectures cannot be quantitatively tested with the fixed-coupling AdS data considered here.
\end{abstract}

\begin{keyword}
ModMax electrodynamics \sep topological AdS black holes \sep extremality \sep charge normalization \sep Weak Gravity Conjecture \sep swampland
\end{keyword}
\end{frontmatter}
\section{Introduction}\label{sec:intro}
The Weak Gravity Conjecture (WGC) is motivated in part by the expectation that charged black holes should be able to discharge, but a classical metric does not determine whether the required charged states exist in the underlying quantum theory. A meaningful comparison requires the charge appearing in the conjecture to be canonically normalized, the candidate state spectrum to be specified, and the decay or extremality criterion to match the asymptotic setting \,\cite{AHMV2007,Harlow2023}. These requirements become particularly important in nonlinear electrodynamics, where the conserved electric charge is a flux of the displacement tensor and need not coincide with the integration constant in the field strength.

ModMax electrodynamics is a conformal, electromagnetic-duality-invariant nonlinear theory parameterized by a dimensionless coupling $\gamma$ \cite{Bandos2020,Kruglov2021,Anand2025}. On a purely electric background, the constitutive relation reduces to a Maxwell-like equation with a $\gamma$-dependent coefficient. This makes the background solution tractable, but it also creates a normalization trap: the field-strength amplitude, the displacement flux, and the charge measured by a canonically rescaled potential must not be assigned the same symbol without a conversion. A charge-to-mass ratio can otherwise acquire a spurious dependence on $\gamma$.

The topological ModMax/ModAMax--AdS family was studied by Eslam Panah, Hamil, and Rodrigues \cite{EslamPanah2026}. While the background solution and its basic thermodynamics were established in \cite{EslamPanah2026}, the present work focuses on the charge normalization, the sensitivity of the extremality curve to the charge convention, the classification of all extremal roots, and the limitations of WGC diagnostics in this setting. A previous study by Gashti and collaborators examined ModMax black holes using horizon, weak-cosmic-censorship, and photon-sphere diagnostics \cite{GashtiModMax2025}. More recently, Gashti, Pourhassan, and Sakall{\i} derived a probe-based WGC criterion for selected asymptotically flat black holes by solving a charged-scalar radial problem and relating its near-horizon conformal weight to a relaxation-time condition \cite{Gashti2026CFT}. That result uses a particular background and a global mode-matching argument. Earlier work by some of the present authors considered WGC diagnostics for charged rotating AdS black holes in a different matter environment \cite{SadeghiGashti2024}; those results likewise do not replace the normalization and spectrum checks required here. The CFT result cannot be transferred to the present topological AdS family merely by copying a near-horizon algebraic factor: the AdS boundary conditions, horizon topology, and complete radial problem must be reconsidered.

A separate strong-field analysis of black-hole decay reports that ModMax fails a necessary charge-growth condition in its asymptotically flat setting \cite{Barbosa2025}. This result is not, by itself, a theorem for every AdS ensemble; it is, however, a reason not to advertise a larger unnormalized charge-to-mass ratio as evidence for WGC compatibility. We therefore keep three questions separate throughout: (i) what follows exactly from the classical extremality curve, (ii) what can be tested locally for an explicitly added probe, and (iii) what remains unknown without a charged spectrum and a controlled ultraviolet completion.
Over the past few years, the study of ModMax black holes has expanded in several directions. For instance, their thermodynamic behavior, photon spheres, and weak cosmic censorship have been examined in various setups \cite{GashtiModMax2025}. Similar thermodynamic analyses, including non-perturbative corrections to entropy and phase transitions, have also been carried out for other black hole solutions in AdS and dirty black hole backgrounds \cite{Soroushfar2024}, and in other gravitational frameworks such as Kalb--Ramond gravity \cite{KalbRamondModMax2025}. Topological black holes in anti-de Sitter space, with spherical, planar, and hyperbolic horizons, also have a long history, starting from the classic work of Birmingham \cite{Birmingham1999}. However, a systematic look at how the charge normalization affects both the extremality curve and the WGC diagnostics in this specific topological ModMax--AdS family has been missing so far.
The present work focuses on those questions and does not attempt to infer a complete stability theorem from thermodynamic data. We reconstruct the charge dictionary from the standard action, derive the first law with one convention throughout, classify every real positive extremal root in the spherical, planar, and hyperbolic sectors, and evaluate the limiting cases that change the physical interpretation. As an independent geometric check, we derive the unstable null circular orbit for the spherical extremal branch without turning it into an observational shadow claim. We then compute the local AdS$_2$ BF index for a minimally coupled scalar used only as a probe. Finally, we state explicitly which swampland conjectures are and are not testable with a fixed ModMax parameter and a negative cosmological constant. The principal conclusions are deliberately limited to the calculations actually performed.

\section{Action, field equations, and charge normalization}\label{sec:model}

We use $G=1$ and the standard ModMax Lagrangian
\begin{equation}\label{eq:MMlag}
\cL_{\ModMax}=\frac14\left[-\cF\cosh\gamma+\sqrt{\cF^2+\cG^2}\sinh\gamma\right],
\end{equation}
where
\begin{equation}
\cF=F_{\mu\nu}F^{\mu\nu},\qquad \cG=F_{\mu\nu}\widetilde F^{\mu\nu},\qquad
\widetilde F^{\mu\nu}=\frac12\epsilon^{\mu\nu\rho\sigma}F_{\rho\sigma},\qquad \gamma\geq0.
\end{equation}
The action is
\begin{equation}\label{eq:action}
I=\frac{1}{16\pi}\int_{\mathcal M}\dd^4x\sqrt{-g}
\left[R-2\Lambda+4\eta\cL_{\ModMax}\right]+I_{\partial\mathcal M},
\qquad \eta=\pm1.
\end{equation}
The sign $\eta=+1$ is the ordinary-sign ModMax branch, while $\eta=-1$ reverses the gauge-sector sign and is referred to as ModAMax. The boundary term depends on the chosen variational problem and does not change the local field equations.

The equations following from this action can be written as
\begin{align}
G_{\mu\nu}+\Lambda g_{\mu\nu}&=2\eta\,\tau_{\mu\nu},\label{eq:einstein}\\
\nabla_\mu\left(\frac{\partial\cL_{\ModMax}}{\partial F_{\mu\nu}}\right)&=0,\label{eq:constitutive}
\end{align}
where the metric-variation tensor is defined unambiguously by
\begin{equation}
\tau_{\mu\nu}\equiv-\frac{2}{\sqrt{-g}}\frac{\delta(\sqrt{-g}\,\cL_{\ModMax})}{\delta g^{\mu\nu}}.
\end{equation}
An overall constant multiplying the constitutive equation is immaterial. For a purely electric configuration, $\cF<0$ and $\cG=0$, so
\begin{equation}\label{eq:electricL}
\left.\cL_{\ModMax}\right|_{\cG=0,\cF<0}=-\frac14e^{\gamma}\cF.
\end{equation}
Thus $e^{+\gamma}$ multiplies the field-strength term. The $e^{-\gamma}$ in the lapse below appears only after rewriting the solution in terms of the conserved displacement charge.

For the static ansatz
\begin{equation}\label{eq:metric}
\dd s^2=-f(r)\dd t^2+\frac{\dd r^2}{f(r)}+r^2\dd\Omega_k^2,
\end{equation}
let $\mathfrak q$ denote the field-strength integration constant and choose
\begin{equation}\label{eq:potentialfield}
A_t(r)=-\frac{\mathfrak q}{r},\qquad F_{rt}=\frac{\mathfrak q}{r^2},\qquad F_{tr}=-\frac{\mathfrak q}{r^2},\qquad \cF=-\frac{2\mathfrak q^2}{r^4}.
\end{equation}
The displacement field is proportional to $e^\gamma\mathfrak q$. For a spherical horizon we define the flux-normalized charge by
\begin{equation}
q\equiv\frac{1}{4\pi}\int_{S^2}e^\gamma\,{}^\star F=e^\gamma\mathfrak q,
\qquad Q=\frac{q}{4\pi},\qquad \mathfrak q=4\pi Qe^{-\gamma},
\label{eq:qdef}
\end{equation}
where the orientation is chosen so that $q\geq0$. Thus the integrated displacement flux is $4\pi q$ in this convention; $Q$ is the per-unit-horizon-volume charge density used below. For planar and hyperbolic horizons, all thermodynamic quantities below are likewise densities per unit horizon volume. The potential difference is therefore
\begin{equation}\label{eq:potentialQ}
A_t(r)=-\frac{q e^{-\gamma}}{r},\qquad U=A_t(\infty)-A_t(r_+)=\frac{q e^{-\gamma}}{r_+}.
\end{equation}
The variables $A_t=-q/r$ and $Q=q/(4\pi)$ would be inconsistent with the lapse term $q^2e^{-\gamma}/r^2$ if used simultaneously; Eqs.~\eqref{eq:qdef}--\eqref{eq:potentialQ} fix this ambiguity.

On this purely electric background it is useful to introduce a Maxwell-normalized field and charge,
\begin{equation}\label{eq:canonicalfield}
\widehat A_\mu=e^{\gamma/2}A_\mu,\qquad \widehat Q=e^{-\gamma/2}Q.
\end{equation}
If a probe has covariant derivative $D_\mu=\nabla_\mu-i e_\Phi A_\mu$, its coupling in these variables is $\widehat e_\Phi=e^{-\gamma/2}e_\Phi$. This field redefinition concerns the purely electric background. The nonanalyticity of ModMax at $\cF=\cG=0$ means that it does not establish equivalence of the full theory, including generic dyonic perturbations, to Maxwell theory \cite{Bandos2020,Barbosa2025}.

\section{Solution and thermodynamics}\label{sec:thermo}

We focus on the AdS branch \(\Lambda=-3/l^2<0\). With the normalization above, the exact pure-electric solution is \cite{EslamPanahModMaxAdS2025,AlBadawi2025,Ahmed2026,Sekhmani2025}
\begin{equation}\label{eq:lapse}
f(r)=k-\frac{m}{r}+\frac{r^2}{l^2}
+\frac{\eta q^2e^{-\gamma}}{r^2}
=k-\frac{m}{r}+\frac{r^2}{l^2}
+\frac{\eta(4\pi Q)^2e^{-\gamma}}{r^2}.
\end{equation}
For \(\eta=+1\) this is the ordinary-sign ModMax branch; for \(\eta=-1\) it is the sign-reversed ModAMax branch. The scalar curvature is \(R=4\Lambda=-12/l^2\), while the Kretschmann invariant is
\begin{equation}\label{eq:Kretschmann}
R_{\mu\nu\rho\sigma}R^{\mu\nu\rho\sigma}
=\frac{24}{l^4}+\frac{12m^2}{r^6}
-\frac{48\eta m q^2e^{-\gamma}}{r^7}
+\frac{56q^4e^{-2\gamma}}{r^8}.
\end{equation}
For nonzero charge, the final term makes the curvature invariant diverge as \(r\to0\). For a neutral solution with \(m\ne0\), the mass term also diverges. The special case \(m=q=0\) must not be described as a curvature singularity solely on the basis of Eq.~\eqref{eq:Kretschmann}. At large radius, \(f(r)\sim r^2/l^2\) and the invariant approaches \(24/l^4\), as required for asymptotically AdS geometry.

Solving \(f(r_+)=0\) gives
\begin{equation}\label{eq:massparam}
m=kr_++\frac{r_+^3}{l^2}+\frac{\eta q^2e^{-\gamma}}{r_+}.
\end{equation}
The Hawking temperature is
\begin{equation}\label{eq:temperature}
T=\frac{f'(r_+)}{4\pi}
=\frac{1}{4\pi}\left(\frac{k}{r_+}+\frac{3r_+}{l^2}
-\frac{\eta q^2e^{-\gamma}}{r_+^3}\right).
\end{equation}
Because the gravitational theory is Einstein gravity, the entropy obeys the area law. Per unit horizon volume, the thermodynamic variables are
\begin{equation}\label{eq:thermovars}
S=\frac{r_+^2}{4},\qquad M=\frac{m}{8\pi},\qquad
Q=\frac{q}{4\pi},\qquad U=\frac{q e^{-\gamma}}{r_+}.
\end{equation}
For a spherical horizon these are densities divided by \(4\pi\); the total quantities are \(S_{\rm tot}=\pi r_+^2\), \(M_{\rm tot}=m/2\), and \(Q_{\rm tot}=q\). This distinction avoids mixing total ADM quantities with densities when comparing different horizon topologies.

In terms of \(r_+\), the mass density is
\begin{equation}\label{eq:massdensity}
M=\frac{1}{8\pi}\left(kr_++\frac{r_+^3}{l^2}
+\frac{\eta q^2e^{-\gamma}}{r_+}\right).
\end{equation}
Equivalently, using \(r_+=2\sqrt S\) and \(q=4\pi Q\),
\begin{equation}\label{eq:MSQ}
M(S,Q;\gamma,\eta)=\frac{k\sqrt S}{4\pi}
+\frac{S^{3/2}}{\pi l^2}
+\eta\frac{\pi Q^2e^{-\gamma}}{\sqrt S}.
\end{equation}
At fixed \(Q\), \(\gamma\), and \(l\), differentiation yields
\begin{align}
\left(\frac{\partial M}{\partial S}\right)_{Q,\gamma}
&=\frac{k}{8\pi\sqrt S}+\frac{3\sqrt S}{2\pi l^2}
-\eta\frac{\pi Q^2e^{-\gamma}}{2S^{3/2}}=T,\label{eq:dMdS}\\
\left(\frac{\partial M}{\partial Q}\right)_{S,\gamma}
&=\eta\frac{2\pi Qe^{-\gamma}}{\sqrt S}=\eta U.
\label{eq:dMdQ}
\end{align}
Thus the non-extended first law is
\begin{equation}\label{eq:firstlaw}
\dd M=T\dd S+\eta U\dd Q.
\end{equation}
The first law is an internal consistency check. It does not independently determine whether the gauge field has a unitary ultraviolet completion or whether the WGC holds.
\begin{figure}[htbp]
  \centering
  \includegraphics[height=7cm,width=10cm]{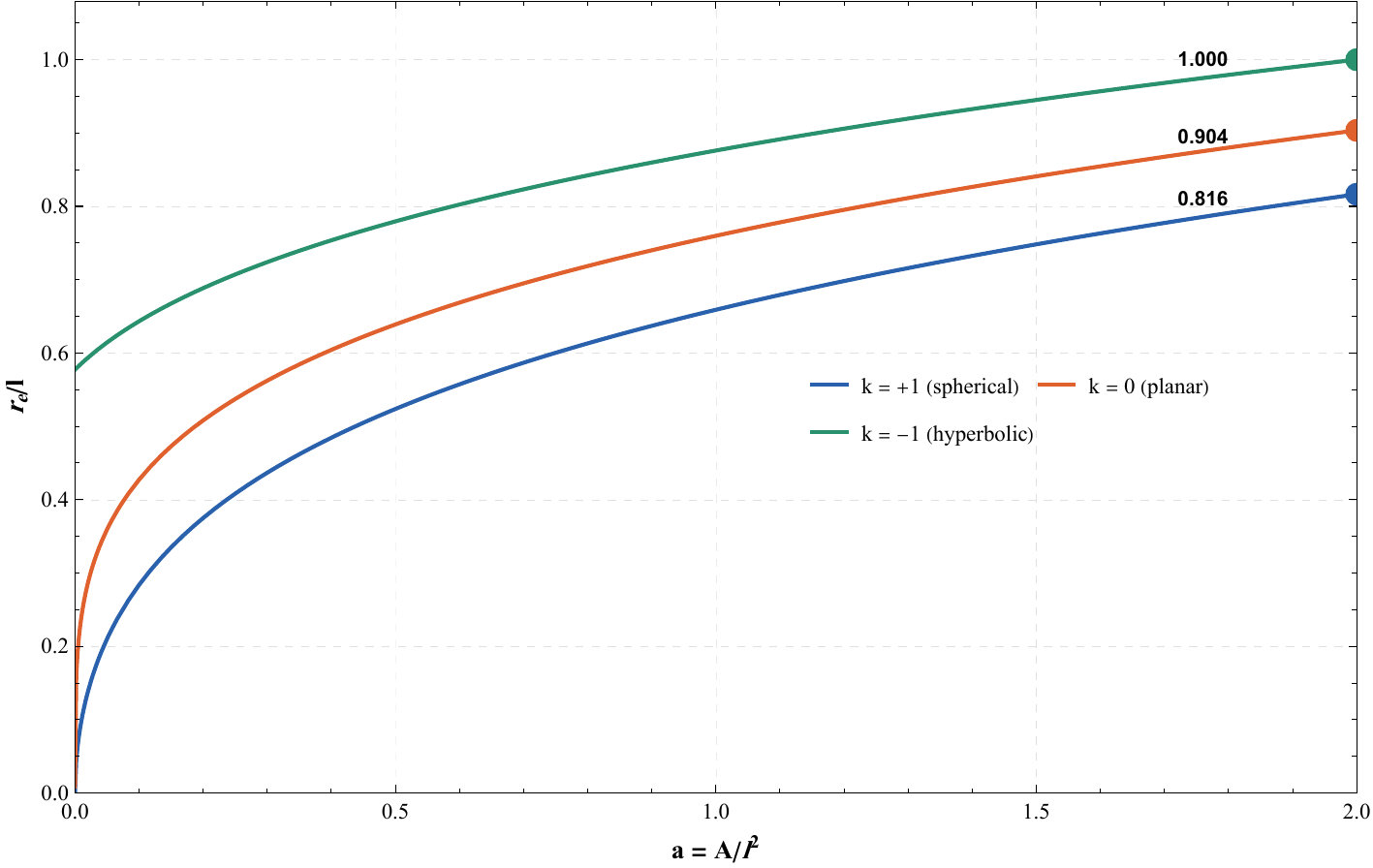}
  \caption{{\small Dimensionless extremal horizon radius $r_e/l$ as a function
of the effective charge parameter
$a=\mathcal A/l^2=16\pi^2Q^2e^{-\gamma}/l^2$ for the ordinary-sign
ModMax branch. The curves correspond to spherical ($k=+1$), planar
($k=0$), and hyperbolic ($k=-1$) horizons. The numerical labels give
the respective values of $r_e/l$ at $a=2$}}
  \label{fig1}
\end{figure}
Figure~\ref{fig1} shows that the extremal radius depends on the horizon topology through the quadratic extremality equation. At fixed effective charge, the hyperbolic branch has the largest extremal radius, whereas the
spherical branch has the smallest one. In the neutral limit, the
spherical and planar extremal radii shrink to zero, while the
hyperbolic branch approaches $r_e/l=1/\sqrt{3}$.
\section{Exact extremality and topology dependence}\label{sec:extremality}

\subsection{General extremal equation}
At a degenerate horizon, \(f(r_e)=f'(r_e)=0\), equivalently \(T(r_e)=0\). Define
\begin{equation}\label{eq:Adef}
x\equiv r_e^2,\qquad u\equiv\frac{x}{l^2},\qquad
\cA\equiv q^2e^{-\gamma}=16\pi^2Q^2e^{-\gamma}\geq0.
\end{equation}
The temperature condition is the quadratic equation
\begin{equation}\label{eq:xquad}
\frac{3x^2}{l^2}+kx-\eta\cA=0,
\end{equation}
with roots
\begin{equation}\label{eq:xroots}
x_\pm=\frac{l^2}{6}\left[-k\pm
\sqrt{k^2+\frac{12\eta\cA}{l^2}}\right].
\end{equation}
Only real, positive roots can represent finite-radius degenerate horizons. The extremal mass simplifies after use of Eq.~\eqref{eq:xquad}:
\begin{equation}\label{eq:mext}
m_{\rm ext}=2r_e\left(k+\frac{2r_e^2}{l^2}\right),
\qquad
M_{\rm ext}=\frac{r_e}{4\pi}\left(k+\frac{2r_e^2}{l^2}\right).
\end{equation}
The sign of \(M_{\rm ext}\) therefore depends on topology and the root, not only on the sign of \(\eta\).

\subsection{ModMax: \texorpdfstring{\(\eta=+1\)}{eta=+1}}
For spherical horizons,
\begin{equation}\label{eq:xplus}
x_e^{(k=+1)}=\frac{l^2}{6}\left[-1+\sqrt{1+\frac{192\pi^2Q^2e^{-\gamma}}{l^2}}\right].
\end{equation}
It is positive for nonzero charge. The formal zero-charge limit has \(x_e\to0\), not a finite-radius extremal Schwarzschild--AdS black hole. For planar horizons,
\begin{equation}\label{eq:xplanar}
x_e^{(k=0)}=\left(\frac{16\pi^2Q^2e^{-\gamma}l^2}{3}\right)^{1/2}.
\end{equation}
This is likewise positive for \(Q\ne0\) and tends to zero when the charge vanishes.

For hyperbolic horizons the positive root is instead
\begin{equation}\label{eq:xhyperMM}
x_e^{(k=-1)}=\frac{l^2}{6}\left[1+\sqrt{1+
\frac{192\pi^2Q^2e^{-\gamma}}{l^2}}\right].
\end{equation}
The discriminant has a plus sign. The alternative root, proportional to \(1-\sqrt{1+192\pi^2Q^2e^{-\gamma}/l^2}\), is negative for nonzero charge and is not a physical radius. Hence the hyperbolic ModMax branch has a real positive extremal radius for every \(Q\), rather than the upper-charge restriction stated in the original draft.

The extremal mass for \(k=-1\) is
\begin{equation}\label{eq:MhyperMM}
M_{\rm ext}^{(k=-1,\eta=+1)}
=\frac{r_e}{4\pi}\left(-1+\frac{2r_e^2}{l^2}\right).
\end{equation}
Since \(r_e^2/l^2\geq1/3\), this mass is negative for \(1/3\leq u<1/2\), zero at \(u=1/2\), and positive for \(u>1/2\). The zero-mass threshold is
\begin{equation}\label{eq:hyperzeromass}
\cA=\frac{l^2}{4},\qquad
Q^2e^{-\gamma}=\frac{l^2}{64\pi^2}.
\end{equation}
Thus it is incorrect to classify all hyperbolic ModMax extremal solutions as negative-mass solutions. A positive-mass WGC comparison is available only above the threshold in Eq.~\eqref{eq:hyperzeromass}; the zero-mass point makes a charge-to-mass ratio singular and is not a standard WGC state.
\begin{figure}[htbp]
  \centering
  \includegraphics[height=7cm,width=10cm]{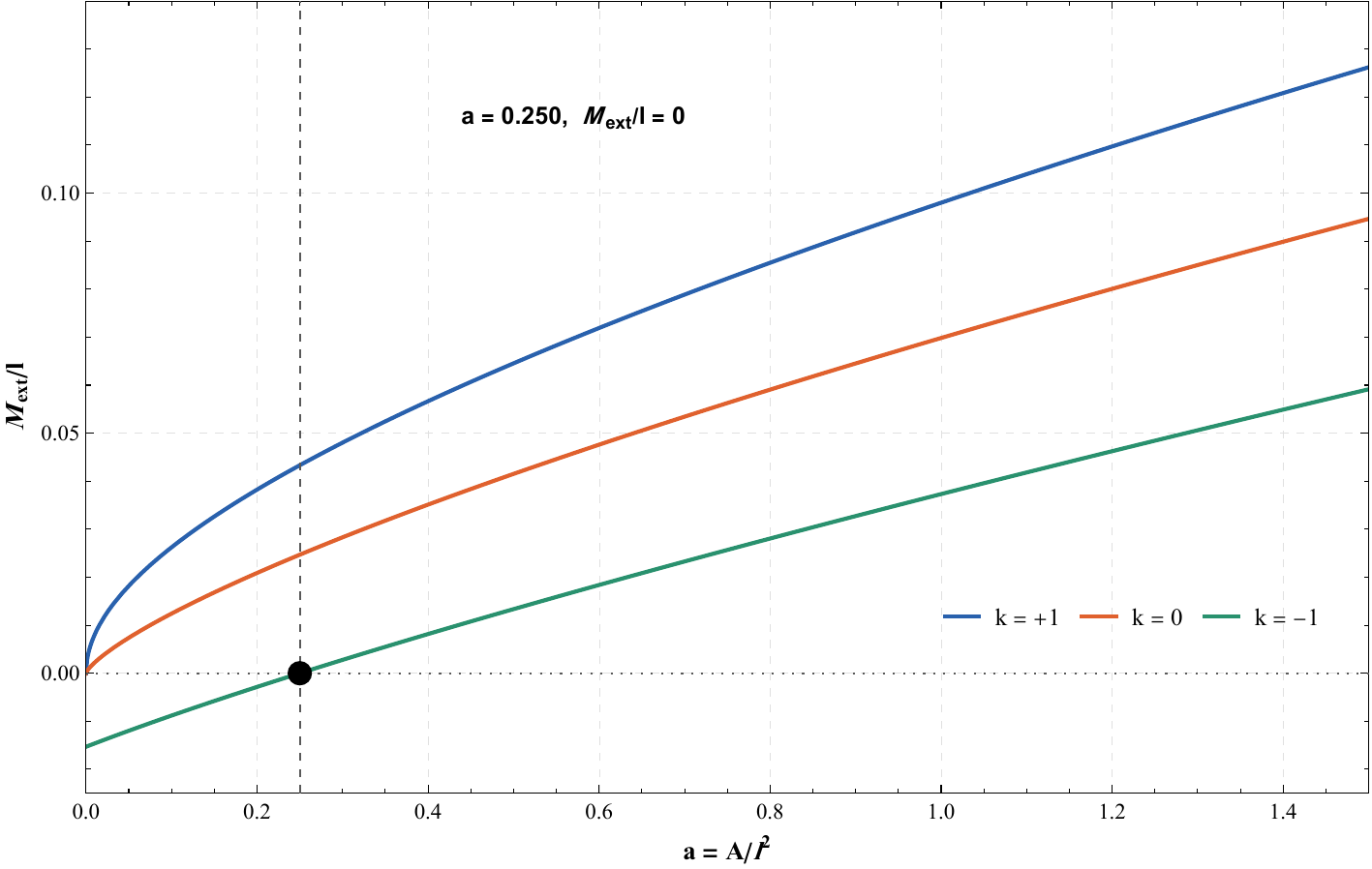}
  \caption{{\small Dimensionless extremal mass density $M_{\rm ext}/l$ versus
the effective charge $a$ for the three topologies of the ordinary-sign
ModMax solution. The horizontal dotted line marks zero mass, and the
vertical dashed line identifies the hyperbolic zero-mass point
$a=1/4$. The spherical and planar branches remain non-negative,
whereas the hyperbolic branch changes from negative to positive mass}}
  \label{fig2}
\end{figure}
Figure~\ref{fig2} shows that the spherical and planar extremal masses are non-negative for positive effective charge. The hyperbolic ModMax branch behaves differently: its extremal mass changes sign at $a=1/4$. This point
corresponds to $r_e^2/l^2=1/2$, where the mass vanishes. A
charge-to-mass ratio is singular there because its denominator
vanishes; this singularity is not, by itself, evidence for a
superextremal state.
Similar ModMax black hole solutions with different horizon topologies and thermodynamic properties have been studied in \cite{EslamPanahModMaxAdS2025,AlBadawi2025}.
\subsection{ModAMax: \texorpdfstring{\(\eta=-1\)}{eta=-1}}
For \(k=+1\) and \(k=0\), Eq.~\eqref{eq:xquad} has no positive solution for \(\cA>0\). Thus these branches have no finite-radius extremal black hole. For the hyperbolic case, real roots exist only if
\begin{equation}\label{eq:AMcondition}
\cA\leq\frac{l^2}{12},\qquad
Q^2e^{-\gamma}\leq\frac{l^2}{192\pi^2},
\end{equation}
and are
\begin{equation}\label{eq:xAM}
x_\pm^{(k=-1,\eta=-1)}=\frac{l^2}{6}
\left[1\pm\sqrt{1-\frac{192\pi^2Q^2e^{-\gamma}}{l^2}}\right].
\end{equation}
For a nonzero charge below the discriminant bound, both algebraic roots are positive. They are not physically interchangeable. Since
\begin{equation}\label{eq:fpp}
f''(r_e)=\frac{2}{r_e^2}\left(k+\frac{6r_e^2}{l^2}\right),
\end{equation}
the smaller root has \(f''(r_e)<0\) and is a local maximum of \(f\), not the outer degenerate horizon. The physical outer extremal branch is the larger root \(x_+\), for which \(f''>0\), except at the endpoint where the two roots coincide. At
\begin{equation}
\cA=\frac{l^2}{12},\qquad r_e^2=\frac{l^2}{6},
\end{equation}
we have \(f''(r_e)=0\); the horizon is more degenerate and the usual \(\AdS_2\) throat approximation below is not valid.

On the physical outer ModAMax extremal branch, \(l^2/6<r_e^2\leq l^2/3\) for nonzero charge. Consequently,
\begin{equation}\label{eq:MhyperAM}
M_{\rm ext}^{(k=-1,\eta=-1)}
=\frac{r_e}{4\pi}\left(-1+\frac{2r_e^2}{l^2}\right)<0.
\end{equation}
The negativity follows from the root range. It does not follow from the expression \(-r_e^3/(4\pi l^2)\), which is not the correct extremal mass formula. In addition, the wrong-sign gauge kinetic sector is a separate obstruction to interpreting ModAMax as a healthy, unitary charged sector without a specified completion.
\begin{figure}[htbp]
  \centering
  \includegraphics[height=7cm,width=10cm]{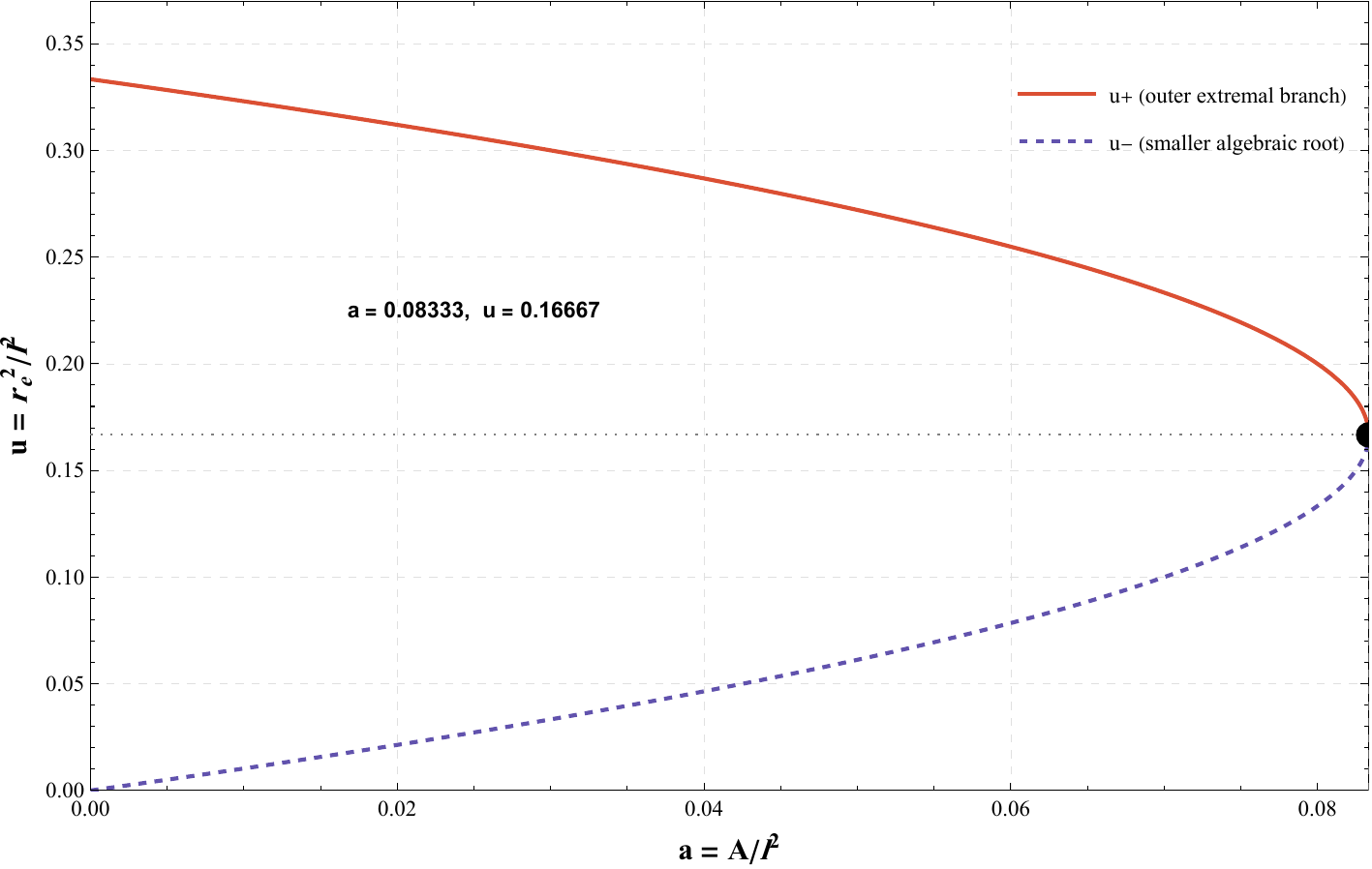}
  \caption{{\small The two algebraic extremality roots
$u_\pm=r_{e,\pm}^2/l^2$ for the hyperbolic ModAMax branch as functions
of $a=\mathcal A/l^2$. The solid curve represents the larger root,
which defines the physical outer extremal branch. The dashed curve
shows the smaller algebraic root. Both merge at the triple-root
endpoint $a=1/12$, $u=1/6$, where the usual quadratic
AdS$_2$-throat approximation breaks down.}}
  \label{fig3}
\end{figure}
The Fig. (\ref{fig3}) shows the two hyperbolic ModAMax roots merge at the discriminant endpoint
$a=1/12$, where $u=1/6$. For nonzero charge below this endpoint,
both algebraic roots are positive, but they do not have the same
geometric status. The smaller root satisfies $f''(r_e)<0$ and is
not the outer degenerate horizon. At the merger, $f''(r_e)=0$, so
the standard quadratic near-horizon expansion leading to an
AdS$_2$ throat cannot be used.
\subsection{Extremal charge-to-mass diagnostics}
For any admissible extremal root with \(M_{\rm ext}\ne0\), Eqs.~\eqref{eq:Adef} and \eqref{eq:mext} give the displacement-charge ratio
\begin{equation}\label{eq:Rraw}
\mathcal R_{\rm disp}\equiv\frac{Q^2}{M_{\rm ext}^2}
=\frac{\eta e^\gamma(k+3u)}{(k+2u)^2},
\qquad u=\frac{r_e^2}{l^2}.
\end{equation}
This formula is useful algebraically but should not be compared with a canonical WGC threshold before the charge convention is fixed. Using the canonical charge \(\widehat Q=e^{-\gamma/2}Q\) gives
\begin{equation}\label{eq:Rcan}
\mathcal R_{\rm can}\equiv\frac{\widehat Q^2}{M_{\rm ext}^2}
=e^{-\gamma}\mathcal R_{\rm disp}
=\frac{\eta(k+3u)}{(k+2u)^2}.
\end{equation}
The explicit factor \(e^\gamma\) has disappeared. If the canonical charge is held fixed, \(u\) is also independent of \(\gamma\) on the purely electric solution. The parameter dependence of Eq.~\eqref{eq:Rraw} at fixed displacement charge therefore cannot be read as a physical enhancement of the WGC without making the same field redefinition for the particle sector.

In the asymptotically flat spherical limit, \(l\to\infty\), the ModMax extremal solution obeys
\begin{equation}\label{eq:flatlimit}
r_e=4\pi |Q|e^{-\gamma/2},\qquad
M_{\rm ext}=|Q|e^{-\gamma/2},\qquad
\frac{|Q|}{M_{\rm ext}}=e^{\gamma/2},\qquad
\frac{|\widehat Q|}{M_{\rm ext}}=1.
\end{equation}
The final equality is the normalization-invariant Reissner--Nordstr\"om value for this pure-electric sector. The preceding ratio differs because it uses displacement charge rather than the canonically normalized charge. This does not make the displacement charge incorrect; it means that both sides of a conjectural inequality must use the same convention.

For the planar branch, the displacement ratio becomes
\begin{equation}\label{eq:planarratio}
\mathcal R_{\rm disp}^{(k=0)}
=\frac{3\sqrt3}{16\pi}\frac{l}{|Q|}
 e^{3\gamma/2},
\end{equation}
whereas the canonical ratio is \(e^{-\gamma}\mathcal R_{\rm disp}\). Its dependence on the charge and AdS scale is another reason that a universal threshold equal to one cannot be imposed on every topological AdS density. In the hyperbolic ModMax branch, Eq.~\eqref{eq:Rraw} can diverge at the zero-mass point, but that divergence is a zero of the denominator, not evidence for a superextremal state.
\begin{figure}[htbp]
  \centering
  \includegraphics[height=8cm,width=15cm]{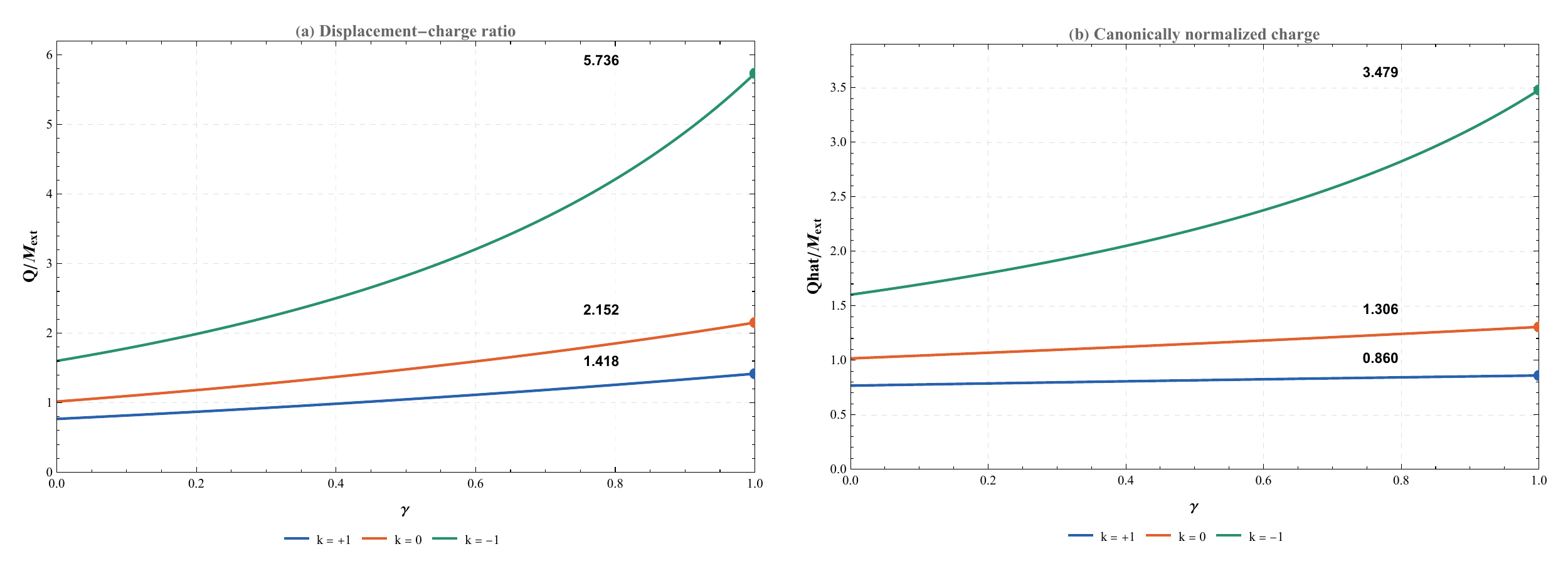}
  \caption{{\small Charge-to-mass ratios along a fixed-displacement-charge
parameter path, with $l=1$ and $Q=0.1$. Panel (a) shows
$Q/M_{\rm ext}$, whereas panel (b) shows
$\widehat Q/M_{\rm ext}$, where
$\widehat Q=e^{-\gamma/2}Q$. The curves correspond to spherical,
planar, and hyperbolic ModMax extremal black holes. The bold
numbers give the ratios at $\gamma=1$. Since the canonical charge
varies along this path, the figure is not a WGC test at fixed
canonical charge}}
  \label{fig4}
\end{figure}
Figure~\ref{fig4} shows that, at fixed displacement charge, the ratio \(Q/M_{\rm ext}\) changes with \(\gamma\) because the metric depends on the combination \(Q^2e^{-\gamma}\). The ratio formed with the rescaled charge,
$\widehat Q=e^{-\gamma/2}Q$, follows a different curve. These
results must be interpreted along the specified parameter path:
holding $Q$ fixed does not hold $\widehat Q$ fixed. At fixed
canonical charge, the purely electric background and its
extremality curve are independent of $\gamma$.
\subsection{Numerical benchmark and limiting cases}
For a numerical benchmark, set $l=1$, $Q=0.1$, and $\eta=+1$. The exact root formulas above give the values in Table~\ref{tab:numerics}. The charge-to-mass entries use both the displacement convention and the Maxwell-normalized charge $\widehat Q=e^{-\gamma/2}Q$. The latter is the appropriate variable for comparison with a canonically normalized Maxwell sector. Neither column is a WGC pass/fail test without a specified particle spectrum and decay criterion.
\begin{table}[t]
\centering
\small
\caption{Numerical extremality benchmark for ModMax with $l=1$ and $Q=0.1$. Masses are the densities defined in Eq.~\eqref{eq:thermovars}. The ratio $\widehat Q/M_{\rm ext}$ uses the rescaled charge; neither ratio alone tests the WGC.}
\label{tab:numerics}
\begin{tabular}{ccccccc}
\toprule
$\gamma$ & $k$ & $x_e=r_e^2$ & $r_e$ & $M_{\rm ext}$ & $Q/M_{\rm ext}$ & $\widehat Q/M_{\rm ext}$\\
\midrule
0.0 & $+1$ & 0.577750 & 0.760099 & 0.130379 & 0.766993 & 0.766993\\
0.0 & 0 & 0.725520 & 0.851774 & 0.098354 & 1.016731 & 1.016731\\
0.0 & $-1$ & 0.911084 & 0.954507 & 0.062450 & 1.601292 & 1.601292\\
\midrule
0.5 & $+1$ & 0.422437 & 0.649951 & 0.095420 & 1.048003 & 0.816186\\
0.5 & 0 & 0.565035 & 0.751688 & 0.067598 & 1.479335 & 1.152107\\
0.5 & $-1$ & 0.755770 & 0.869350 & 0.035389 & 2.825761 & 2.200705\\
\midrule
1.0 & $+1$ & 0.303888 & 0.551260 & 0.070530 & 1.417841 & 0.859964\\
1.0 & 0 & 0.440050 & 0.663363 & 0.046459 & 2.152419 & 1.305508\\
1.0 & $-1$ & 0.637221 & 0.798261 & 0.017434 & 5.736049 & 3.479090\\
\bottomrule
\end{tabular}
\end{table}
At fixed displacement charge, increasing $\gamma$ decreases $\cA=16\pi^2Q^2e^{-\gamma}$ and shrinks the extremal radius. The raw ratio can consequently grow. At fixed canonical charge $\widehat Q$, the lapse is independent of $\gamma$ in the pure-electric sector, so this change is a normalization-path effect rather than an independent weakening of gravity.

The limiting values are equally important because formal charge-to-mass ratios can become singular at zero mass or where the extremal radius shrinks to zero. Table~\ref{tab:limits} records the directly evaluated cases. Values of $M/l$ use the same density normalization as above.
\begin{table}[t]
\centering
\small
\caption{Analytic limiting cases of the extremal branches. The dimensionless variable is $u=r_e^2/l^2$ and $a=\cA/l^2$.}
\label{tab:limits}
\begin{tabularx}{\textwidth}{@{}>{\raggedright\arraybackslash}p{0.28\textwidth}ccccY@{}}
\toprule
Branch / limit & $a$ & $u$ & $r_e/l$ & $M_{\rm ext}/l$ & Interpretation\\
\midrule
ModMax, $k=+1$, $Q\to0$ & $0$ & $0$ & $0$ & $0$ & No finite-radius neutral extremal horizon.\\
ModMax, $k=0$, $Q\to0$ & $0$ & $0$ & $0$ & $0$ & The charged extremal radius shrinks to zero.\\
ModMax, $k=-1$, $Q\to0$ & $0$ & $1/3$ & $0.577350$ & $-0.015309$ & Neutral hyperbolic extremal limit.\\
ModMax, $k=-1$, $M_{\rm ext}=0$ & $1/4$ & $1/2$ & $0.707107$ & $0$ & $Q^2e^{-\gamma}=l^2/(64\pi^2)$; $Q/M$ is singular.\\
ModAMax, $k=-1$, triple root & $1/12$ & $1/6$ & $0.408248$ & $-0.021651$ & $f''(r_e)=0$; the usual quadratic AdS$_2$ throat is absent.\\
\bottomrule
\end{tabularx}
\end{table}

For another explicit check, take ModAMax with $l=1$, $Q=0.01$, and $\gamma=0.5$. Then $a=\cA/l^2\simeq0.00957795<1/12$, and the roots are $u_-=0.00987021$ and $u_+=0.32346312$. The smaller root has $f''<0$ and is not an outer extremal horizon. The larger root gives $r_e\simeq0.568738$, $M_{\rm ext}\simeq-0.0159797$, and $L_2^2\simeq0.343825$. These values follow from the exact root and mass formulas; they are not fits to observational data.


\subsection{Thermodynamic character of the zero-mass point on the hyperbolic ModMax branch}
\label{sec:zero-mass}

The zero-mass point identified in Eq.~\eqref{eq:hyperzeromass} deserves a separate thermodynamic reading. The charge-to-mass diagnostic of Eq.~\eqref{eq:Rraw} diverges there, while the horizon radius and entropy density remain finite and nonzero. We show that this divergence is a pole of the ratio, not a signal of a superextremal state, and that the zero-mass point is a smooth zero crossing of the extremal mass along the connected extremal branch.

It is convenient to write the extremal mass~\eqref{eq:MhyperMM} and the entropy density~\eqref{eq:thermovars} in terms of \(u = r_e^2/l^2\). For \(r_e = l\sqrt{u}\), they take the compact forms
\begin{equation}
	M_{\mathrm{ext}}(u) = \frac{l}{4\pi}\,\sqrt{u}\,(2u-1),
	\qquad
	S_{\mathrm{ext}}(u) = \frac{l^2}{4}\,u .
	\label{eq:Mext-Sext-u}
\end{equation}
The zero of the mass occurs at \(u = 1/2\), i.e.\ \(r_e^2 = l^2/2\), in agreement with Eq.~\eqref{eq:hyperzeromass}. At that point
\begin{equation}
	S_{\mathrm{ext}}\!\left(\tfrac{1}{2}\right) = \frac{l^2}{8} \neq 0,
	\qquad
	T_{\mathrm{ext}} = 0,
	\label{eq:zero-mass-S}
\end{equation}
so the vanishing of the extremal mass does not coincide with a vanishing horizon radius or a vanishing entropy density. The Bekenstein--Hawking area law gives a finite, nonzero entropy density.

The derivative of the extremal mass with respect to \(u\),
\begin{equation}
	\frac{\mathrm{d}M_{\mathrm{ext}}}{\mathrm{d}u}
	= \frac{l}{4\pi}\,\frac{6u-1}{2\sqrt{u}},
	\label{eq:dMdu}
\end{equation}
is finite and positive at \(u = 1/2\):
\begin{equation}
	\left.\frac{\mathrm{d}M_{\mathrm{ext}}}{\mathrm{d}u}\right|_{u=1/2}
	= \frac{l}{4\pi}\sqrt{2} \;>\; 0 .
	\label{eq:dMdu-half}
\end{equation}
Thus \(M_{\mathrm{ext}}\) crosses zero with a nonvanishing, finite slope. The zero is a simple root of the extremal mass function, not a degenerate point of the mass function itself. Equivalently, starting from the general mass function \(M(S,Q)\), the response to the displacement charge is
\begin{equation}
	\left(\frac{\partial M}{\partial Q}\right)_{\!S,\gamma}
	= \eta\,\frac{4\pi Q\,e^{-\gamma}}{r_+},
\end{equation}
which at the extremal point \(r_+ = r_e\) and \(u = 1/2\) evaluates to
\begin{equation}
	\left(\frac{\partial M}{\partial Q}\right)_{\!S,\gamma}
	\;\xrightarrow[u=1/2]{}\;
	\eta\,\frac{\sqrt{2}}{2}\,e^{-\gamma/2}.
	\label{eq:dMdQ-half}
\end{equation}
This value is finite and nonzero at the zero-mass point, whereas the ratio $Q/M_{\mathrm{ext}}$ diverges. The divergence of the charge-to-mass diagnostic at \(u = 1/2\) is therefore entirely due to the vanishing of its denominator $M_{\mathrm{ext}}$. The numerator and the thermodynamic response of the mass remain regular.

\begin{figure}[ht!]
	\centering
	\begin{minipage}[t]{0.48\textwidth}
		\centering
		\includegraphics[width=\textwidth]{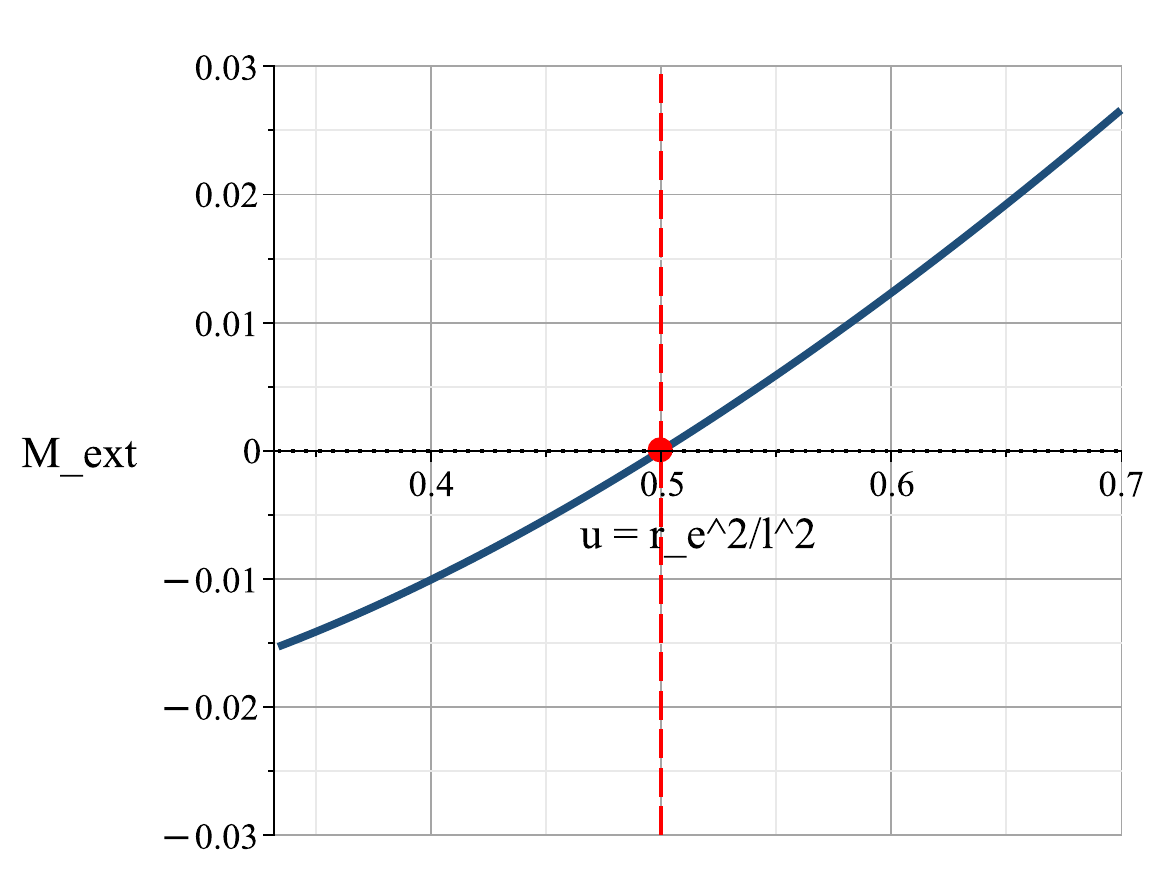}
		\\[4pt]
		\small (a)
	\end{minipage}\hfill
	\begin{minipage}[t]{0.48\textwidth}
		\centering
		\includegraphics[width=\textwidth]{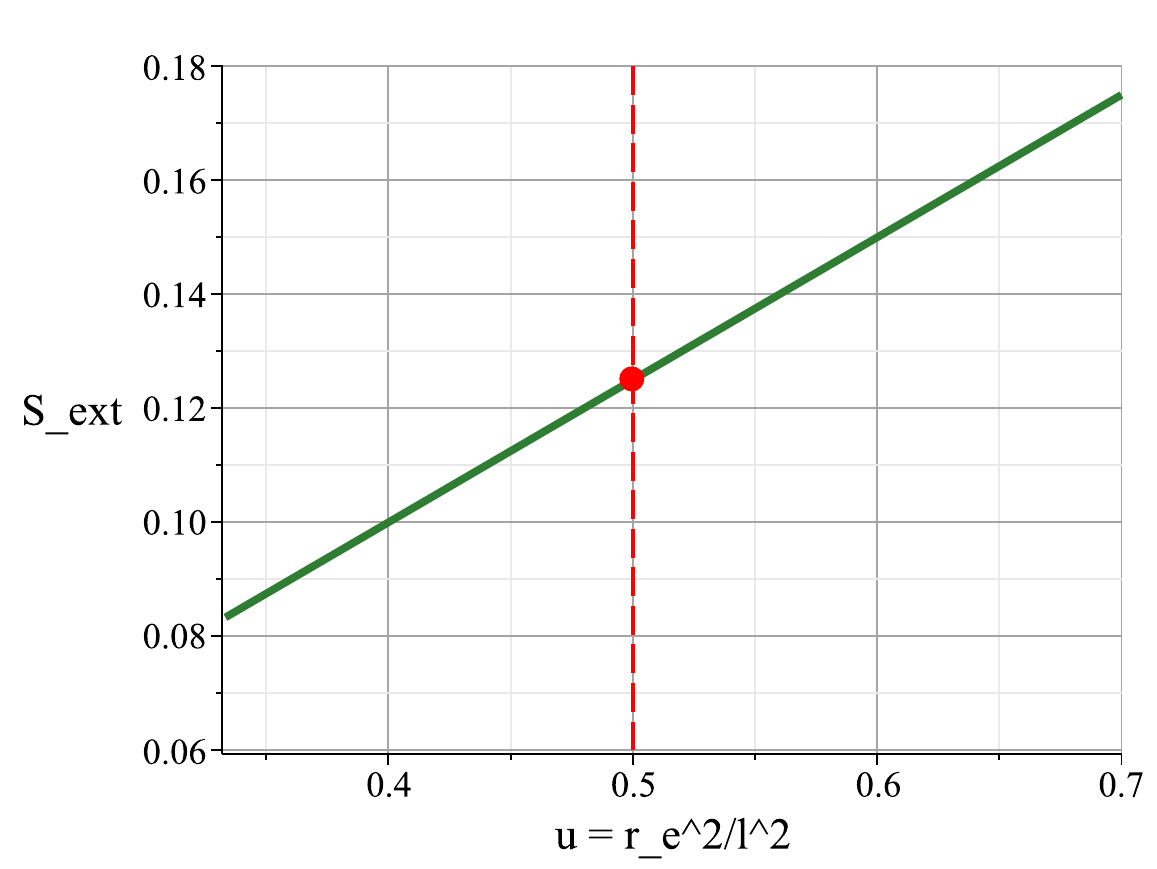} 		\\[4pt]
		\small (b)
	\end{minipage}
	
	\vspace{12pt} 
	
	\begin{minipage}[t]{0.48\textwidth}
		\centering
		\includegraphics[width=\textwidth]{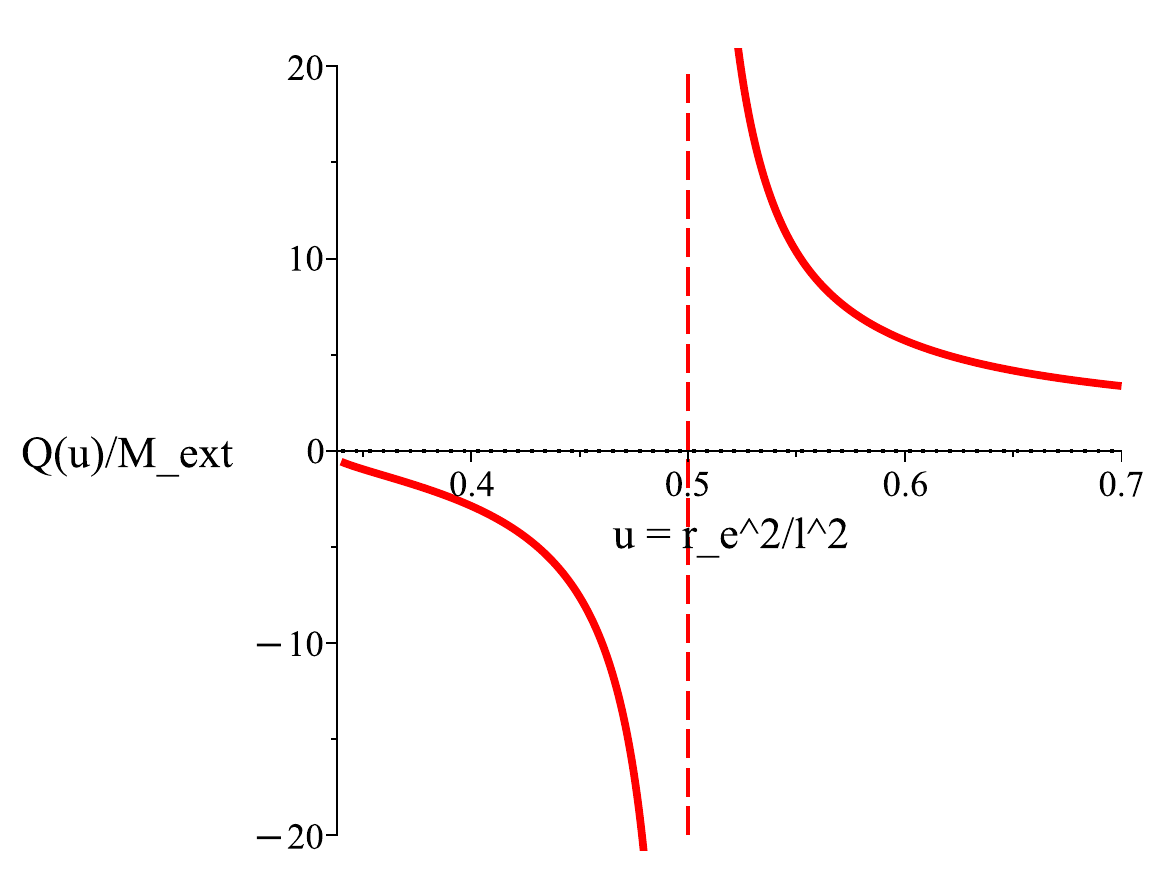} 		\\[4pt]
		\small (c)
	\end{minipage}
	
	\caption{Hyperbolic ModMax extremal branch near the zero-mass point (\(l=1\), \(\gamma=0.5\), \(Q = Q(u) = e^{\gamma/2}\sqrt{u(3u-1)}/(4\pi)\) along the extremal branch). (a) The extremal mass crosses zero at \(u=1/2\) with finite slope [Eq.~\eqref{eq:dMdu-half}]. (b) The entropy density stays finite and smooth across \(u=1/2\) [Eq.~\eqref{eq:zero-mass-S}]. (c) The ratio \(Q(u)/M_{\mathrm{ext}}\) diverges at \(u=1/2\) because the denominator vanishes. The branch is a single connected family with a smooth zero crossing of the extremal mass.}
	\label{fig:zero-mass}
\end{figure}

Figure~\ref{fig:zero-mass} summarizes these results visually: Fig~\ref{fig:zero-mass} (a) shows the finite-slope zero crossing of the extremal mass, Fig~\ref{fig:zero-mass} (b) shows the continuous and monotonic entropy density, and Fig~\ref{fig:zero-mass} (c) shows the pole in the ratio \(Q/M_{\mathrm{ext}}\). Table~\ref{tab:zero-mass} records the corresponding numerical behaviour for \(l = 1\) and \(\gamma = 0.5\). The entropy density varies smoothly and monotonically across \(u = 1/2\), the extremal mass changes sign, and only the ratio \(Q/M_{\mathrm{ext}}\) exhibits a pole. The negative- and positive-mass segments belong to the same connected extremal family, and the zero-mass point has a finite horizon radius and entropy density.

\begin{table}[t]
	\centering
	\caption{Thermodynamic quantities on the hyperbolic ModMax extremal branch near the zero-mass point, for \(l = 1\) and \(\gamma = 0.5\). The displacement charge \(Q(u)\) is evaluated along the extremal branch. The entropy density is smooth across \(u = 1/2\), the extremal mass changes sign, and the charge-to-mass ratio has a pole at \(u = 1/2\) only because \(M_{\mathrm{ext}}\) vanishes there. For the negative-mass segment, the signed ratio $Q/M_{\mathrm{ext}}$ is reported only as an algebraic diagnostic. It is not interpreted as a physical WGC charge-to-mass ratio. The conventional comparison is restricted to states with positive mass and a consistently
		normalized charge.}
	\label{tab:zero-mass}
	\begin{tabular}{cccccc}
		\hline\hline
		$u$ & $r_e$ & $M_{\mathrm{ext}}$ & $S_{\mathrm{ext}}$ & $Q(u)$ & $Q/M_{\mathrm{ext}}$ \\
		\hline
		0.40 & 0.6325 & $-0.0101$ & 0.1000 & 0.0289 & $-2.87$ \\
		0.45 & 0.6708 & $-0.0053$ & 0.1125 & 0.0406 & $-7.60$ \\
		0.49 & 0.7000 & $-0.0011$ & 0.1225 & 0.0490 & $-44.0$ \\
		0.50 & 0.7071 & $ 0.0000$ & 0.1250 & 0.0511 & pole \\
		0.51 & 0.7141 & $+0.0011$ & 0.1275 & 0.0531 & $+46.7$ \\
		0.55 & 0.7416 & $+0.0059$ & 0.1375 & 0.0611 & $+10.4$ \\
		0.60 & 0.7746 & $+0.0123$ & 0.1500 & 0.0708 & $+5.74$ \\
		\hline\hline
	\end{tabular}
\end{table}

Two points are worth emphasizing. First, the zero-mass point is not a WGC state: a charge-to-mass comparison requires a nonvanishing mass, and the divergence of $Q/M_{\mathrm{ext}}$ at \(u = 1/2\) comes from the denominator, not from a superextremal ratio. This is in line with the warning in Sec.~\ref{sec:swampland} that a raw charge-to-mass ratio is not a WGC test~\cite{AHMV2007,Harlow2023,Palti2019}. Second, the zero-mass point is a smooth zero crossing of the extremal mass along the connected extremal branch, with finite horizon radius and entropy density. The entropy is continuous and monotonic, the temperature vanishes along the branch, and the mass crosses zero with finite slope. A similar picture emerges in other ModMax black hole backgrounds, where thermodynamic transitions and singular ratios have been studied in \cite{AlBadawi2026,EslamPanahRainbow2024}.
We stress that these results hold for the fixed-coupling, purely electric background considered throughout. They do not address the spectrum, the ultraviolet completion, or the global perturbation problem~\cite{BF1982}, and they do not change the conclusion of Sec.~\ref{sec:swampland}.

\FloatBarrier
\section{Spherical null circular orbit as a geometric check}\label{sec:photon}

For the spherical sector ($k=+1$), the null-geodesic effective potential per unit squared angular momentum is $V_{\rm null}(r)=f(r)/r^2$. A circular null orbit satisfies $V_{\rm null}'(r_{\rm ph})=0$, or
\begin{equation}\label{eq:photoncondition}
r_{\rm ph} f'(r_{\rm ph})-2f(r_{\rm ph})=0.
\end{equation}
For the ordinary-sign ModMax lapse this reduces exactly to
\begin{equation}\label{eq:photonroots}
2r_{\rm ph}^2-3m r_{\rm ph}+4\cA=0,
\qquad r_{\rm ph}^{\pm}=\frac{3m\pm\sqrt{9m^2-32\cA}}{4},
\end{equation}
provided the discriminant is non-negative. The cosmological term cancels in Eq.~\eqref{eq:photoncondition}, although the AdS scale remains in $m$ and $\cA$ along an extremal family. The outer circular orbit corresponds to the plus root. At a circular orbit, the second derivative of $V_{\rm null}$ is
\begin{equation}
V_{\rm null}''(r_{\rm ph})=\frac{3m-4r_{\rm ph}}{r_{\rm ph}^5}=-\frac{\sqrt{9m^2-32\cA}}{r_{\rm ph}^5}<0
\end{equation}
for the outer root with nonzero discriminant, so that orbit is unstable in the radial direction.

On the spherical extremal ModMax branch, $m=2r_e(1+2u)$ and $\cA=r_e^2(1+3u)$, where $u=r_e^2/l^2$. Substitution into Eq.~\eqref{eq:photonroots} gives the simple exact relation
\begin{equation}\label{eq:photonext}
r_{\rm ph}=2r_e(1+3u),\qquad \frac{r_{\rm ph}}{r_e}=2(1+3u)>2.
\end{equation}
Thus the outer null circular orbit lies outside the extremal horizon. Null geodesics and photon spheres in ModMax black holes have also been examined in \cite{Pantig2022}. For the numerical values in Table~\ref{tab:numerics}, the resulting radii are shown in Table~\ref{tab:photon}. These radii are geometric diagnostics; an observational shadow prediction would additionally require an observer prescription and a physical environment, and is not claimed here.
\begin{table}[t]
\centering
\small
\caption{Outer null circular-orbit radius for the extremal spherical ModMax branch at $l=1$ and $Q=0.1$. Values are obtained from Eq.~\eqref{eq:photonext}.}
\label{tab:photon}
\begin{tabular}{cccc}
\toprule
$\gamma$ & $r_e$ & $r_{\rm ph}$ & $r_{\rm ph}/r_e$\\
\midrule
0.0 & 0.760099 & 4.155082 & 5.466502\\
0.5 & 0.649951 & 2.947282 & 4.534620\\
1.0 & 0.551260 & 2.107650 & 3.823328\\
\bottomrule
\end{tabular}
\end{table}
At fixed displacement charge the radius decreases as $\gamma$ increases in this benchmark because the effective geometric charge $\cA$ decreases. At fixed canonical charge, $\cA=(4\pi\widehat Q)^2$ is unchanged, and so is this purely electric geometry. This orbit calculation is independent of the WGC verdict and cannot replace a charged-spectrum or decay analysis.

\section{Near-horizon geometry and a charged-probe test}\label{sec:probe}

A double root with $f''(r_e)>0$ has a local near-horizon geometry $\mathrm{AdS}_2\times\Sigma_k$. Expanding the lapse gives
\begin{equation}
f(r)=\frac12f''(r_e)(r-r_e)^2+\mathcal O((r-r_e)^3),
\end{equation}
with
\begin{equation}\label{eq:fppvalue}
f''(r_e)=2\left(\frac{k}{r_e^2}+\frac{6}{l^2}\right),\qquad
L_2^2=\frac{2}{f''(r_e)}=\frac{r_e^2}{k+6r_e^2/l^2}.
\end{equation}
This formula applies only when $k+6r_e^2/l^2>0$. It applies to the ordinary ModMax extremal branches with a positive finite radius and to the physical outer ModAMax hyperbolic root away from the triple-root endpoint. At that endpoint $f''=0$ and this quadratic AdS$_2$ approximation fails.

The electric field at the horizon, in the gauge convention of Eq.~\eqref{eq:potentialQ}, is
\begin{equation}\label{eq:Efield}
\mathcal E_h=|F_{rt}(r_e)|=\frac{|q|e^{-\gamma}}{r_e^2}=\frac{4\pi|Q|e^{-\gamma}}{r_e^2},\qquad e_d=\mathcal E_h L_2^2.
\end{equation}
For the ModMax branch, extremality gives the useful identity
\begin{equation}\label{eq:EL2}
\mathcal E_h^2 L_2^2=e^{-\gamma}\frac{k+3u}{k+6u},\qquad u=\frac{r_e^2}{l^2},
\end{equation}
where the ratio is evaluated on the regular branch for which $L_2^2>0$.

To test a specific local stability condition, add a minimally coupled complex scalar as a probe; it is not part of the background action:
\begin{equation}\label{eq:probeaction}
I_\Phi=-\int\dd^4x\sqrt{-g}\left[|D\Phi|^2+\mu_\Phi^2|\Phi|^2\right],\qquad D_\mu=\nabla_\mu-i e_\Phi A_\mu.
\end{equation}
Let $\lambda_k$ be an eigenvalue of the negative Laplacian on the unit horizon section. On a sphere $\lambda_k=\ell(\ell+1)$; on a planar section it is the squared wave number, and for a compact hyperbolic quotient it belongs to the scalar spectrum of that quotient. The near-horizon index is
\begin{equation}\label{eq:nusquared}
\nu_{\lambda_k}^2=\frac14+L_2^2\left(\mu_\Phi^2+\frac{\lambda_k}{r_e^2}\right)-e_\Phi^2e_d^2.
\end{equation}
The local charged-scalar Breitenlohner--Freedman condition is
\begin{equation}\label{eq:BF}
\left[\mu_\Phi^2+\frac{\lambda_k}{r_e^2}-e_\Phi^2\mathcal E_h^2L_2^2\right]L_2^2\geq-\frac14,
\qquad \nu_{\lambda_k}^2\geq0.
\end{equation}
A violation indicates a local BF instability of this probe in the near-horizon throat. Satisfaction of Eq.~\eqref{eq:BF} is not a proof of full stability: it does not solve the global radial problem with AdS boundary conditions, coupled metric--gauge perturbations, or possible superradiant modes \cite{BF1982}. The choice $\lambda_k=0$ below is a local constant-mode benchmark; global normalizability on noncompact planar sections, and the spectrum on compact horizon quotients, must be imposed separately.

As a reproducible numerical check, set $l=1$, $Q=0.1$, $\gamma=0.5$, $\lambda_k=0$, $\mu_\Phi=0.2$, and $e_\Phi=1$. Table~\ref{tab:probe} gives the horizon data and BF index calculated from Eqs.~\eqref{eq:fppvalue}--\eqref{eq:nusquared}. The critical probe coupling is defined by $\nu_0^2=0$:
\begin{equation}\label{eq:ecrit}
e_{\Phi,\BF}=\frac{\sqrt{1/4+\mu_\Phi^2L_2^2}}{\mathcal E_hL_2^2}.
\end{equation}
\begin{table}[t]
\centering
\small
\caption{Local charged-probe test on the three ModMax extremal branches at $l=1$, $Q=0.1$, and $\gamma=0.5$. The illustrative probe has $\mu_\Phi=0.2$, $e_\Phi=1$, and $\lambda_k=0$. Positive $\nu_0^2$ means that this probe does not violate the local AdS$_2$ BF bound for these parameters; it does not establish global stability.}
\label{tab:probe}
\begin{tabular}{ccccc}
\toprule
$k$ & $L_2^2$ & $\mathcal E_h$ & $e_{\Phi,\BF}$ & $\nu_0^2$\\
\midrule
$+1$ & 0.119514 & 1.804268 & 2.340794 & 0.208282\\
$0$ & 0.166667 & 1.348923 & 2.253455 & 0.206122\\
$-1$ & 0.213819 & 1.008493 & 2.358059 & 0.212054\\
\bottomrule
\end{tabular}
\end{table}
These numbers are an internal consistency check for a deliberately chosen probe, not a prediction for a known particle. They do not supply a WGC bound because $e_\Phi$ and $\mu_\Phi$ are inputs rather than data derived from the spectrum of the theory.

\section{Parameter sensitivity and the meaning of fixed charge}\label{sec:response}

The parameter $\gamma$ appears in the metric only through $Q^2e^{-\gamma}$ when the displacement-charge convention of Eq.~\eqref{eq:qdef} is used. Rewriting Eq.~\eqref{eq:MSQ} in terms of the Maxwell-normalized charge $\widehat Q=e^{-\gamma/2}Q$ gives
\begin{equation}\label{eq:Mcanonical}
M(S,\widehat Q;\eta)=\frac{k\sqrt S}{4\pi}+\frac{S^{3/2}}{\pi l^2}+\eta\frac{\pi\widehat Q^2}{\sqrt S}.
\end{equation}
The canonical mass function has no explicit $\gamma$-dependence in this purely electric sector. At fixed $S$ and $Q$, however, the chain rule yields
\begin{equation}\label{eq:gamma-chain}
\left(\frac{\partial M}{\partial\gamma}\right)_{S,Q}
=-\frac{Q}{2}\left(\frac{\partial M}{\partial Q}\right)_{S,\gamma}
=-\eta\frac{\pi Q^2e^{-\gamma}}{\sqrt S},
\end{equation}
whereas
\begin{equation}\label{eq:gamma-canonical}
\left(\frac{\partial M}{\partial\gamma}\right)_{S,\widehat Q}=0.
\end{equation}
The same statement holds for a smooth extremal branch because $\partial M/\partial S=T=0$ at extremality. Thus the derivative at fixed $Q$ measures the change in the canonical charge along that path; it is not an invariant measure of a physical correction to the extremality bound.

This observation is deliberately restricted to the pure-electric background. It does not say that the full nonlinear constitutive law, generic dyonic solutions, or quantum fluctuations are independent of $\gamma$. It also shows why a variation of $\gamma$ at fixed noncanonical charge should not be presented as an independent WGC test.

\section{What can be concluded about swampland conjectures?}\label{sec:swampland}

\subsection{Electric WGC}
The exact extremality curve is one necessary ingredient in a black-hole WGC discussion, but it is not sufficient to establish the conjecture. A particle-level comparison would require at least: (i) the normalization of the compact gauge group and its canonical coupling, (ii) the masses and charges of the allowed states, and (iii) a decay or discharge criterion compatible with the topology, ensemble, and AdS boundary conditions. None of these spectral or global ingredients follows from the static metric.

Two recent results are relevant but do not settle the present AdS problem. Gashti and collaborators studied ModMax with horizon, photon-sphere, and weak-cosmic-censorship diagnostics \cite{GashtiModMax2025}; such diagnostics are useful characterizations of a background but do not by themselves identify a superextremal particle. Gashti, Pourhassan, and Sakall{\i} later derived a probe/CFT criterion for selected asymptotically flat black holes, including a purely electric Einstein--ModMax branch, by solving the charged-scalar radial problem and imposing a relaxation-time condition \cite{Gashti2026CFT}. The latter is a global argument in its stated setting; only the local BF calculation is repeated here. We do not import that CFT threshold into topological AdS without solving the corresponding radial problem and boundary-value problem. Other aspects of ModMax black holes, such as their scattering properties and rotating solutions, have also been explored in \cite{Baptista2025,Bixano2026}.

A separate strong-field study reports that ModMax fails a necessary charge-growth condition for black-hole decay in an asymptotically flat setting \cite{Barbosa2025}. Its result should not be asserted as a theorem for every AdS topology, but it prevents a blanket claim that the ModMax theory is WGC-compatible solely because some raw charge-to-mass ratios exceed unity. Our conclusions are therefore intentionally limited: the classical extremality diagnostic is established; a WGC pass/fail verdict is not.

When it comes to the distance and de Sitter conjectures \cite{OoguriVafa2007,Obied2018}, things are even more out of reach in our current setup. These conjectures are deeply tied to the global structure of the moduli space and the sign of scalar potentials, which means they demand information about the ultraviolet completion that a purely electric, fixed-\(\gamma\) background simply cannot offer. As pointed out in recent reviews \cite{vanRiet2023}, testing such ideas usually requires either a controlled string compactification or a well-defined asymptotic boundary problem—and neither of those is available here.

\subsection{The sign-reversed branch}
For $\eta=-1$, the gauge-sector action has the opposite sign. The corresponding Maxwell-like kinetic term around the purely electric branch has negative energy relative to the ordinary-sign theory. The classical solutions may still be analyzed mathematically, but they do not provide a conventional unitary WGC test bed without a separate completion that resolves the wrong-sign sector. The existence of a formal degenerate horizon does not remove this issue. For related studies on the thermodynamic and optical properties of ModMax black holes in various backgrounds, see \cite{Shahzad2024,Siahaan2024,Khosravani2024}.

\FloatBarrier
\section{Conclusions}\label{sec:conclusion}
We have re-examined four-dimensional topological AdS black holes sourced by ModMax electrodynamics and by its sign-reversed counterpart, consistently distinguishing the field-strength amplitude from the conserved displacement charge. With the standard ModMax Lagrangian, the pure-electric field-strength term is proportional to $e^{+\gamma}$; the $e^{-\gamma}$ factor in the metric appears after the solution is expressed in terms of displacement charge. This convention fixes the potential and yields the first law with the same variables throughout.

The extremality equation is quadratic in $x=r_e^2$. Its correct roots show that the hyperbolic ModMax branch has no upper-charge discriminant bound. Its extremal mass is negative for $1/3\leq r_e^2/l^2<1/2$, vanishes at $Q^2e^{-\gamma}=l^2/(64\pi^2)$, and is positive above that threshold. For ModAMax, nonzero-charge extremality is absent for spherical and planar horizons; the hyperbolic outer branch exists only below its discriminant bound and has negative mass. At the endpoint, the root is more degenerate and the usual quadratic AdS$_2$ throat expansion does not apply.
We have also examined the zero-mass point on the hyperbolic extremal branch. Although the charge-to-mass ratio diverges at this point, the horizon radius and entropy density remain finite and nonzero, while the extremal mass crosses zero smoothly. This shows that the divergence of the charge-to-mass ratio results from the vanishing denominator and, by itself, does not indicate a superextremal configuration or provide evidence for the weak gravity conjecture.
For the pure-electric background, the rescaled charge $\widehat Q=e^{-\gamma/2}Q$ removes the explicit $\gamma$-dependence of the Reissner--Nordstr\"om-like extremality curve. The apparent growth of $Q/M_{\rm ext}$ at fixed displacement charge is consequently not, on its own, a normalization-independent strengthening of the WGC. We also evaluated the local charged-scalar BF index for an explicitly stated probe. The chosen parameter set satisfies the local bound on all three ModMax topologies, but this numerical example does not establish stability of the full system.

The supported conclusion is narrower than a claim of WGC consistency: the classical extremality curve and the local probe BF condition are derived, while a genuine WGC test still requires a canonical quantum charge normalization, a charged spectrum, and the relevant black-hole decay calculation. The sign-reversed ModAMax sector has a wrong-sign gauge kinetic term and is not a healthy WGC candidate without a completion. The distance conjectures and the de Sitter conjecture cannot be quantified from a fixed coupling and a negative cosmological constant alone. No claim of full perturbative stability or of cosmic censorship is made.

\end{document}